\documentclass[runningheads,a4paper]{llncs}

\usepackage{amssymb,amsmath}
\usepackage{graphicx}

\usepackage{url}
\urldef{\mailsa}\path|{alfred.hofmann, ursula.barth, ingrid.haas, frank.holzwarth,|
\urldef{\mailsb}\path|anna.kramer, leonie.kunz, christine.reiss, nicole.sator,|
\urldef{\mailsc}\path|erika.siebert-cole, peter.strasser, lncs}@springer.com|    
\newcommand{\keywords}[1]{\par\addvspace\baselineskip
\noindent\keywordname\enspace\ignorespaces#1}

\usepackage{algorithmic} 
\usepackage{algorithm} 
\usepackage{multirow} 
\usepackage{amsfonts}
\usepackage{wrapfig}

\begin{document}

\mainmatter  

\title{Qualitative Modelling  via Constraint Programming: Past, Present and Future}

\titlerunning{Qualitative Modelling  via Constraint Programming}

%
%
\author{Thomas W. Kelsey\inst{1}%
\and Lars Kotthoff\inst{1}
\and Christoffer A.  Jefferson\inst{1}
\and Stephen A. Linton\inst{1}
\and Ian Miguel\inst{1}
\and Peter Nightingale\inst{1}
\and Ian P. Gent\inst{1}}
\authorrunning{Qualitative Modelling  via Constraint Programming}

\institute{
School of Computer Science,\\
University of St Andrews, KY16 9SX, UK\\
}

%
%

\toctitle{Lecture Notes in Computer Science}
\tocauthor{Authors' Instructions}
\maketitle

\begin{abstract}
Qualitative modelling is a technique integrating the fields of theoretical computer science, artificial intelligence and the physical and biological sciences.
The aim is to be able to model the behaviour of systems without estimating
parameter values and fixing the exact quantitative dynamics. Traditional applications are the
study of the dynamics of physical and biological systems at a higher level of abstraction than that obtained by estimation of numerical parameter values for a fixed quantitative model. Qualitative modelling has been studied and implemented to varying degrees of sophistication in Petri nets, process calculi and constraint programming. In this paper we reflect on the strengths and weaknesses of existing frameworks, we demonstrate how recent advances in constraint programming can be leveraged to produce high quality qualitative models, and we describe the advances in theory and technology that would be needed to make constraint programming the best option for scientific investigation in the broadest sense.
\keywords{Constraint Programming, Qualitative Models, Compartmental Models, Dynamical Systems}
\end{abstract}

\section{Introduction}
\label{intro}

The standard approach for non-computer scientists when investigating dynamic scientific systems is to develop a quantitative mathematical model. Differential equations are chosen in the belief that they best represent (for example) convection-diffusion-reaction or population change, and parameter values are estimated from empirical data. This approach suffers from several limitations which are widely documented, and which we summarise with examples in Section \ref{quant}. 

In a standard modelling text \cite[Chapter 5]{Haefner2005}, qualitative model formulation is described as
\begin{quote}
{\it 
$\ldots$ the conversion of an objective statement and a set of hypotheses and assumptions into an informal, conceptual model. This form does not contain explicit equations, but its purpose is to provide enough detail and structure so that a consistent set of equations can be written. The qualitative model does not uniquely determine the equations, but does indicate the minimal mathematical components needed. The purpose of a qualitative model is to provide a conceptual frame-work for the attainment of the objectives. The framework summarizes the modeler's current thinking concerning the number and identity of necessary system components (objects) and the relationships among them.
}
\end{quote}

 For the computer scientist, a qualitative approach is more natural. The dynamics of the system under investigation are described in a formal language, but with no (or few) {\it a priori} assumptions made about the specific mathematical model that may be produced. This means working at a higher level of abstraction than usual, it requires the formalisation of complex system behaviour, and it involves searching a large space of candidate models for those to be used to generate numerical models. Computer scientists are, in general, trained to be able to identify and work at the most suitable levels of abstraction; they also design and use highly formal languages, and routinely develop algorithms for NP-hard  problem classes. Hence the computer scientist is ideally qualified to undertake qualitative modelling. This is by no means a new observation, and in Section \ref{qualold} we give a critical evaluation of historic and current computer science approaches to this problem. We focus on three particular approaches, constraint programming (CP), temporal logics and process calculi. In our view, historic CP approaches were hindered by both struggles to accommodate temporality into constraints, and by limitations in the CP languages and tools available at the time. The process calculus and temporal logic approaches have been more successful: the languages and tools used to model and verify computer system behaviour have been (and are being) adapted to model important systems arising in molecular and cell biology. 
 
 The CP approach has been recently revisited, using languages and tools developed as part of the Constraint Solver Synthesiser research project at St Andrews. We give a detailed worked example in Section \ref{qualnew} in which the application area is human cell population dynamics. A version of this example will be presented at the forthcoming Workshop on Constraint Based Methods for Bioinformatics~\cite{KL2012}. We demonstrate the ability to 
 \begin{enumerate}
 \item describe sophisticated qualitative dynamic behaviour in a non-temporal modelling language;
 \item convert these descriptions into standard CP constraints;
 \item explore the large solution spaces of the resulting constraint satisfaction problems (CSPs);
 \item  iterate using parameter estimates and/or subsidiary modelling assumptions to converge on useable quantitative models.
 \end{enumerate}
 
 However, fundamental problems remain. In particular, our exploration of solution spaces is neither truly stochastic nor targeted enough to reduce non-useful search effort. Nor do we have any organised way to investigate the tradeoff between realism of qualitative model and computational complexity of quantitative model. We explore these and other limitations in Section \ref{future}, and present them as research opportunities for the CP community. Successful research activity would be beneficial to the scientific community in the widest sense. Any scientific team would be able to describe the system under investigation in terms of qualitative system descriptions such as:
\begin{itemize}
 \item behaviour A is required;
 \item behaviour B is forbidden;
 \item if C happens, it happens after D;
 \item the rate of change of the rate of change of E has exactly two minima in timescale F;
 \item the rate of change in the decline of G is no less than the rate of change in the increase in H. 
 \end{itemize} 
 
 CP technology would then be used to iteratively converge on suitable models for use by the global scientific community. 
 In our opinion, this would represent an important transfer of CP expertise, languages and search to our colleagues working in other scientific fields.

\section{Quantitative mathematical models}
\label{quant}  

Successful computer modelling in the physical, biological and economic sciences is a difficult undertaking. Domains are often poorly measured due to ethical, technical and/or financial constraints. In extreme instances the collection of accurate longitudinal data is simply impossible using current techniques.  This adversely affects  the production and assessment of hypothetical quantitative models, since the incompleteness of the domain datas necessitates the making of assumptions that may or may not reflect ground truths. A second category of assumptions are involved in the choice of quantitative modelling framework. Hypothetical solutions can be ruled out by restricting the complexity of models, and unrealistic models can be allowed by over-complex models. For both types of {\it a priori} assumption, mutually exclusive assumptions must be kept  separate, sometimes with no scientific justification. 

The remainder of this section consists of two illustrative examples, both taken from biology. 

\subsection{Nitric Oxide diffusion}
\label{NO}

Our first example (adapted from a paper by Degasperi and Calder presented at a workshop on  Process Algebra and Stochastically Timed Activities~\cite{calder}) of the limitations of starting the modelling process by selecting a mathematical model involves
modelling nitric oxide (NO) bioavailability in blood vessels. Models of this scenario
aim to determine the diffusion distance of NO along the radius of a vessel, where NO is
produced in a narrow region on the internal wall of the vessel. Numerous models have been
developed over the last decade and most share underlying
assumptions and use the similar diffusion governing equations. In particular, a vessel is
modelled as a cylinder with partial differential equations (PDEs), using Fick's law of diffusion
in cylindrical coordinates. Compartments define areas such as endothelium (where NO is produced),
vascular wall, and lumen (i.e. where the blood flows). Another common assumption
is that the diffusion operates only in the radial direction, while it can be considered negligible
in other directions.  A complete review and critical evaluation of these models is given in \cite{Tsoukias2008}. 
The author concludes:

\begin{quote}
{\it
The complexity of NO interactions  in vivo makes
detailed quantitative analyses through mathematical
modeling an invaluable tool in investigations of NO
pathophysiology. Mathematical models can provide
a different perspective on the mechanisms that
regulate NO signaling and transport and can be
utilized for the validation and screening of proposed
hypotheses. At this point, however, the predictive
ability of these models is limited by the lack of
quantitative information for major parameters that
affect NO's fate in the vascular wall. Further, the
difficulties associated with measuring NO directly in
biological tissues and the scarcity of NO measurements
in the microcirculation present a significant
obstacle in model validation. Thus, caution is
needed in interpreting the in silico simulations and
accepting model predictions when experimental data
are missing. {\bf Advances} in both the experimental
methodologies and {\bf in the theoretical models} are
required to further elucidate NO's roles in the
vasculature.}
 \cite[our emphases]{Tsoukias2008}
\end{quote}

\begin{figure}[!ht]
\begin{center}
\includegraphics[width=4.5in]{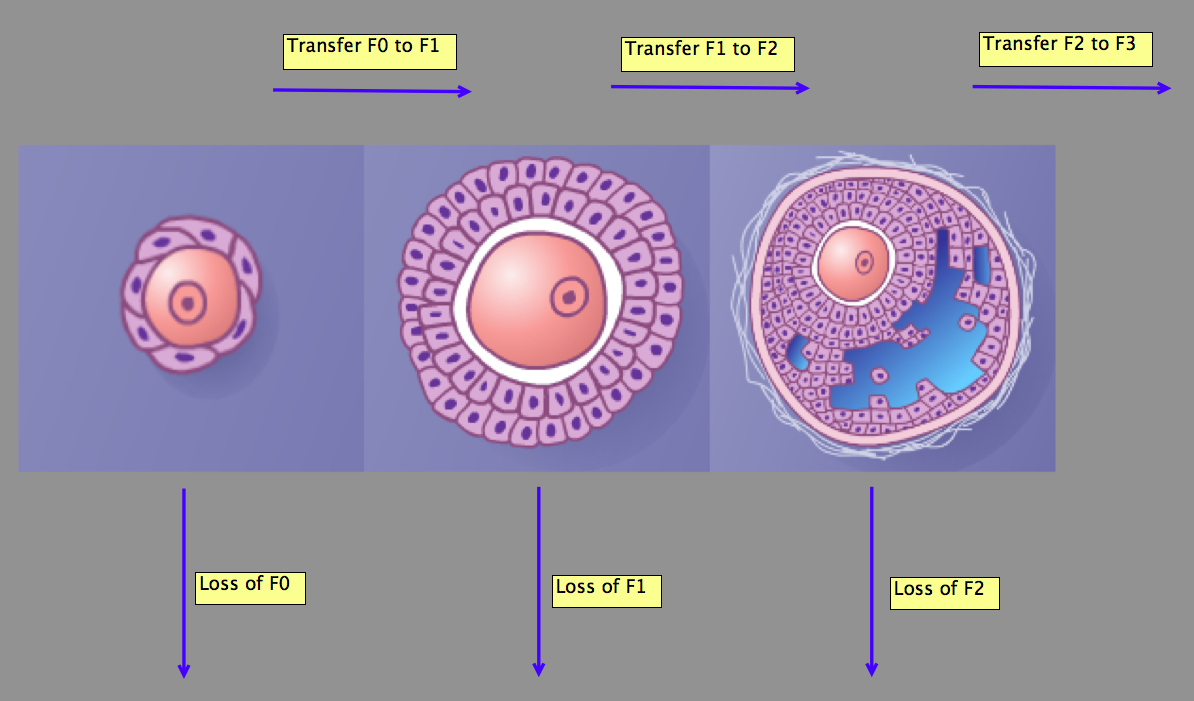}
\end{center}
\caption{Compartmental schematic of human ovarian follicular development.}
\label{follicles}
\end{figure}

\subsection{Ovarian follicle dynamics}
\label{NGF}  

Our second example involves the modelling of human cell populations.
The human ovary contains a population of primordial (or non-growing) follicles (F0 in Figure \ref{follicles}). 
Some of these are recruited towards maturation and start to grow. Many of these die off through atresia, but some become primary follicles (F1 in Figure \ref{follicles}). Again, a proportion of these die off with the remainder growing into secondary follicles (F2 in Figure \ref{follicles}). This continues until a very small proportion become eggs that are released from the ovary for potential fertilisation. For the purposes of this study, we consider only the dynamics of follicle progression (primordial to primary to secondary). Since there are well-defined physiological differences between the types, the obvious choice of quantitative model is compartmental:

$$
\frac{dF_0}{dt} = -k_{T_0}F_0 - k_{L_0}F_0 
$$
$$
\frac{dF_1}{dt} = k_{T_0}F_0 - k_{T_1}F_1 - k_{L_1}F_1 
$$
$$
\frac{dF_2}{dt} = k_{T_1}F_1 - k_{T_2}F_2 - k_{L_2}F_2 
$$

Kinetic loss and transfer parameters --  $k_{L_i}$ and $k_{T_i}$ respectively -- are found in principle by estimating populations at known ages, then fitting ODE solutions that minimise residual errors~\cite{Faddy1995a}. 

There are several limitations to this approach. Empirical data is scarce for primordial follicles~\cite{Wallace2010}, is calculated by inference for primary follicles~\cite{Kelsey2011a}, and simply does not exist for secondary follicles. Mouse-model  studies have produced reasonable parameter estimates and validation~\cite{Bristol-Gould2006}, but it is not known how well these results translate to humans.

As a direct result of these limitations, two entirely different compartmental models have been published in the literature. In \cite{Bristol-Gould2006} there are no losses after $F_0$, whereas in \cite{Faddy1995a} there are no losses for $F_1$, losses for $F_2$, and losses for $F_0$ after age 38. A third research group investigating the same cell dynamics but with its own empirical data and modelling assumptions would be highly likely to produce a third quantitative model being fundamentally different to those already published. So there is an obvious problem: which (if any) of these models should be used by the wider research community to describe and account for changes in cell populations over time?

\begin{figure}[ht]
\begin{minipage}[b]{0.5\linewidth}
\centering
\includegraphics[width=\textwidth]{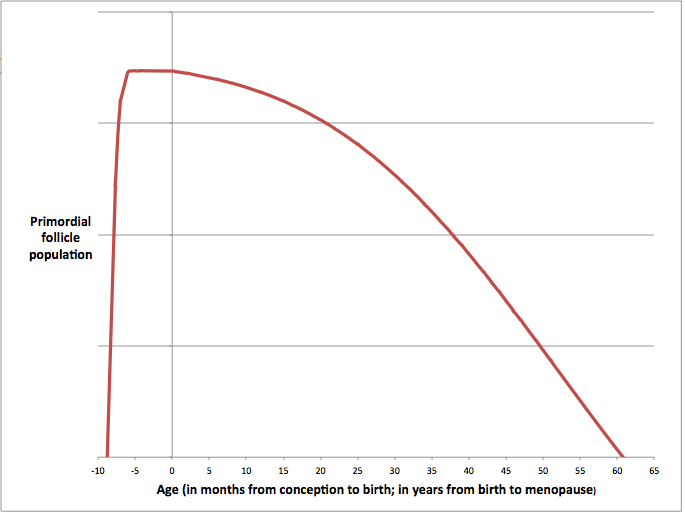}
\end{minipage}
\begin{minipage}[b]{0.5\linewidth}
\centering
\includegraphics[width=\textwidth]{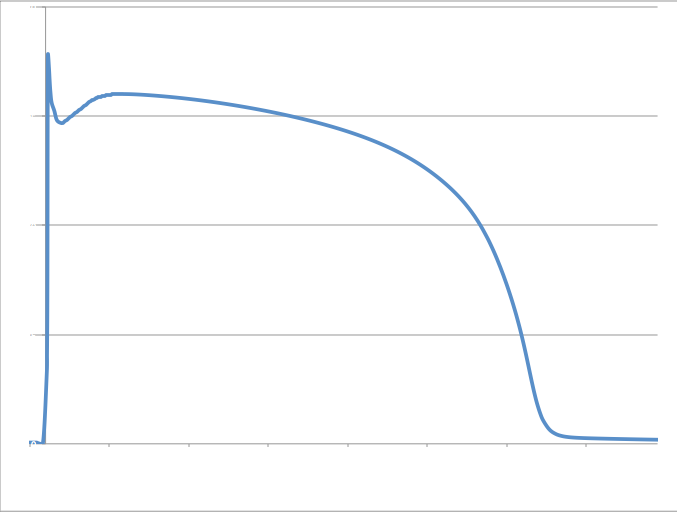}
\end{minipage}
\caption{Two hypothetical models of primordial follicle population from conception to menopause. On the left, a peak model adapted from~\cite{Wallace2010}. On the right, the solution of a reaction--diffusion equation. Both are supported by existing physiological theory and empirical evidence.}
\label{models}
\end{figure}

A more fundamental problem is that the loss--migration model may not be the correct choice. 
Recent studies have shown that human ovarian stem-cells exist, suggesting that further model parameters are needed to allow for regeneration of the primordial follicle pool. The resulting models suffer from biological implausibility in the mouse model~\cite{Bristol-Gould2006}, and remain to be produced for humans. A key methodological  drawback is that the use of compartmental models leads to a constrained class of solutions that excludes other plausible models. For example, the dynamics could also be modelled by nonlinear reaction--diffusion equations that lead to solutions that are unlikely to be obtained from a system of coupled linear ODEs (Figure \ref{models}). 

\section{Existing approaches to QM}
\label{qualold}  

Qualitative modelling is a mature computer scientific technique, with existing methods and results for qualitative compartmental models~\cite{Menzies1992,Menzies1997,Radke-Sharpe1998} and for the use of CSPs to describe and solve qualitative models~\cite{Clancy1998,Escrig2002}. However, these latter studies either reported incomplete algorithms~\cite{Clancy1998} or described complicated algebras with no associated CSP modelling language or optimised CSP solver~\cite{Escrig2002}. In 2002, a hybrid approach was presented in which concurrency was described in terms of CP constraints~\cite{Bockmayr02}.

A key observation is that these studies were published 10--20 years ago. It appears that the limitations of CP technology at the time were collectively sufficient to stifle the development of languages, solvers and tools for  CP-based qualitative modelling.

Other approaches include process calculi and temporal logics, both of  which have been shown to be successful at the molecular level \cite{CH09} and the protein network level \cite{CCFS06,RBFS11}, but not as yet at inter- and intra-cellular levels. Despite this, the process calculus and temporal logic communities are engaging in active current research to improve its techniques and widen access to other scientific areas. Of particular note are BIOCHAM (temporal logic) and BioPEPA (process calculus).

BIOCHAM~\cite{Calzone2006} consists of two languages (one rule-based, the other based on either the CTL or LTL temporal logic languages) that allows the iterative development of quantitative models from qualitative ones. This answers the obvious question posed by newcomers to qualitative modelling: ``given a good qualitative model, how do I derive a model that I can use for numeric studies?" BIOCHAM has sophisticated tool support and is under active current development (version 3.3 released in October 2011). 

BioPEPA~\cite{biopepa} a process algebra for the modelling and the analysis of biochemical networks. It is a modification of PEPA (originally defined for the performance analysis of computer systems), in order to handle the use of general kinetic laws.  The Edinburgh-based BioPEPA research group has sought and received substantial funding to improve the accessibility of their framework by researchers at all levels of systems biology. A cloud-based architecture is under development, as is improved translation to and from SBML (System
Biology Markup Language) formats, thereby supporting easier exchange and curation of models.  

In summary, from the competing candidates for a computer science basis for successful qualitative modelling, CP has -- as it were -- fallen by the wayside, while temporal logics and process calculi are providing real technology and support, at least to the biomedical modelling communities. We see no obvious reason for this: clearly time is a variable in all dynamical modelling, and therefore notions of ``liveness", ``before" and ``after" needed to be incorporated into the qualitative modelling framework. But this is perfectly possible in CP, as we demonstrate in Section \ref{qualnew}.

\section{Case study: cell dynamics QM using constraints}
\label{qualnew}  

Our case study is the compartmental modelling of NGFs described in Section \ref{NGF}.
We use the Savile Row tool that converts constraint problem models formulated in the solver-independent
modelling language Essence$'$~\cite{Frisch2008} to the input format of the Constraint Satisfaction Problem (CSP) solver Minion~\cite{GJM06}. Savile Row converts Essence$'$ problem instances into Minion format and applies reformulations (such as common
subexpression elimination) that enhance search.  As well as the standard variables and constraints expected of a CSP modelling language, Essence$'$ allows the specification of ``for all" and ``exist" constraints, that are then re-cast as basic logic constraints in Minion. 

We expect our candidate qualitative models to be implemented as differential equations or by non-linear curve-fitting. In both case we need to specify the notions of rate of change and smoothness. 
Suppose that $X[0,\ldots,n]$ is a series of variables representing a follicle population at different ages. Then we can approximate first derivatives by $X'[1,\ldots,n]$ where $X'[i] = X[i] - X[i-1]$, and second derivatives by $X''[1,\ldots,n-1]$ where $X''[j] = X'[j+1] - X'[j]$. These definitions allow us to post qualitative constraints about peak populations
$$
\exists p \in [1,\ldots,n] \mathrm{~such~that~}\forall i > p, X'[i] < 0 \wedge \forall i < p, X'[i] > 0 .
$$
We can require or forbid smoothness by restricting the absolute value of the $X''$ variables, and require or forbid fast rates of population growth by restrictions on the $X'[i]$. 

By having three sets of variables (primordial, primary and secondary follicles) each with up to two derivative approximations, we can model interactions between the populations at different ages. For example, we can require a zero population of secondary follicles until puberty, after which the population behaviour is similar to that of primary follicles, but on  a smaller scale and with an adjustable time-lag. 

\begin{table}[ht]
\begin{center}
\begin{tabular}{l|c}  

Essence$'$ statement & Qualitative description \\
\hline
find $x$ :    $[int(0..max)]$ of $int(0..100)$ & percentage of peak population\\
find $y$:  $[int(1..max)]$ of $int(-r\cdots r) $ & 1st deriv. variables \\
find $z$ :   $int(1..max-1)]$ of  $int(-r\cdots r) $ & 2nd deriv. variables \\
forAll $i : int(1..max) .  y[i] = x[i] - x[i-1] $ & 1st deriv.definition \\
forAll $j : int(1..max-1) .  z[j] = y[j+1] - y[j] $ & 2st deriv. definition \\
exists $k,j :  int(2..birth) .  $  & \\
  \quad      forall $i : int(birth..max) . $ & \\
     \qquad      $i < k  \Rightarrow  y[i] > 0 $ & positive 1st deriv. pre-peak \\
        \qquad   $i > k \Rightarrow y[i] < 0$  &  negative  1st deriv. post-peak  \\
           \qquad   $x[k] = 100 \wedge y[k] = 0$   & it is a peak \\
          \qquad    $i > birth  \Rightarrow |z[i]| < max$  & smooth post-gestation \\

\hline
\end{tabular}
\end{center}
\caption{An example of a simple qualitative model specified in Essence$'$. When supplied with values for $max$, $r$,  and $birth$, Savile Row will construct a Minion instance, the solutions of which are all hypothetical models that respect the qualitative description. }
\label{example}
\end{table}

To further abstract away from quantitative behaviour,  populations can be defined in terms of proportion of peak rather than absolute numbers of cells, different time scales can be used for different age ranges (e.g. neonatal vs post-menopausal), and we can model the qualitative behaviour of values that are normally log-adjusted in quantitative studies. Table \ref{example} gives an illustrative example of a model involving one type of follicle. 

Any solution of such a model is a candidate for the basis of a quantitative model of actual cell dynamics, once boundary conditions and scale conditions are supplied. For example, the population of each type of follicle is known to be zero at conception, and can be assumed to be below 1,000 at menopause. Several studies have reported that peak primordial population is about 300,000 per ovary~\cite{Wallace2010}, and there is initial evidence that primary follicle population peaks at 13--15 years of age in humans~\cite{Kelsey2011a}. Using a combination of facts and quantitative information, a range of quantitative models can be produced for later empirical validation.   

Each of our qualitative models represents a class of CSPs, a set of variables with integer or Boolean domains together with a set of constraints involving those variables. A solution is an assignment of domain values to variables such that no constraint is violated. In our methods, solutions are found by Minion using  backtrack search with a variety of search heuristics. 
In general, there will be many more solutions to the CSP than realistic models, and many more realistic models than models that accurately describe reflect what happens in nature. Moreover, the resulting quantitative models can be graded by their complexity --  linear ODE, piecewise-linear ODE, quadratic ODE, ..., non-linear PDE.  Hence the ideal situation would be a CSP solution leading to an easily solved quantitive model that is biologically accurate. However, no such solution need exist, and we need to investigate the tradeoff between model complexity and model accuracy.

We can sample the space of CSP solutions by randomly ordering the variables before making value assignments, thereby constructing a different but logically equivalent search tree at each attempt.  This allows us to estimate the likelihood of ``good" models being found (i.e cheap and accurate), and thereby estimate the computational costs involved in attempting to find the best model that can be derived from our qualitative descriptions. 

In this case study we have utilised recent advances in CSP technology such as solver-independent modelling frameworks, specification--solver interfaces that enhance CSP instances, and the use of solvers that can quickly find all solutions to large and complex CSP instances~\cite{KLR04,Distler2009,DKKJ12}. Taken together, these advances allow us to easily specify qualitative behaviour of cell dynamics, obtain solutions that generate quantitative models, and systematically investigate the tradeoffs between computational expense, model complexity and biological accuracy in a domain for which there is extremely limited direct empirical data. Our investigations utilise the search heuristics used to find CSP solutions: solvers proceed by backtrack search in a tree constructed by explicit choices for current search variable and current value assignment, by randomising these choices we can explore the space of candidate solutions.

The framework for ovarian cells treats primordial follicles as a source, and the other types as both sinks and sources. There is no feedback in the dynamical system, but we see no reason why this aspect could not be included if required. Moreover, it is relatively simple to incorporate other indicators of ovarian reserve~\cite{KWNAW11,KW12,FKAWN12} thereby obtaining an integrated model involving cells, hormones and physiology. We therefore believe that this initial  study can generalise to other domains at other levels of systems biology from population-based epidemiology to steered molecular dynamics.

\section{Future directions for CP}
\label{future}  

The case study in Section \ref{qualnew} was realised using languages and tools developed in the Constraint Solver Synthesiser project at St Andrews. 
Currently, applying constraint technology to a large, complex problem requires significant manual tuning by an expert. Such experts are rare. The central aim of the project is to improve dramatically the scalability of constraint technology, while simultaneously removing its reliance on manual tuning by an expert. It is our view that here are many techniques in the literature that, although effective in a limited number of cases, are not suitable for general use. Hence, they are omitted from current general solvers and remain relatively undeveloped. QM is an excellent example.  

Recent advances in CP technology allow us to
\begin{enumerate}
\item describe complex qualitative system behaviour in a language accessible and understandable by anyone with a reasonable level of scientific and/or mathematical training;
\item  optimise the definition of CSPs based on qualitative descriptions via analysis of the options for variables, values and constraints;
\item use machine-learning to build an optimised  bespoke solver for the class of CSPs derived from the descriptions;
\item efficiently search the solution spaces of large and complex CSP instances.
\end{enumerate}

However, we are at the proof-of-concept stage for QM, having shown the ability in principle to produce useful results, rather than extensive and peer-reviewed research output. We now present specific avenues of research that would allow not only the production of high quality qualitative models, but also a robust schema for deriving a suitable quantitative model from the space of solutions of a CSP that represents a QM. The research areas are given in order of realisability: the first version of Savile Row (Section \ref{srow}) was released in July 2012 and is under current active development, whereas the systematic search for models that are both realistic and lead to computationally inexpensive differential equations (Section \ref{space2}) is a completely unexplored research topic. 

Several of the references for the research topics mentioned in the remainder of this section are incomplete. This is due to the work being part of unfinished investigations, or being planned and designed as future investigations.

\subsection{Essence$'$ and Savile Row}
\label{srow}

Savile Row~\cite{srow} is a 
modelling assistant tool that reads the language Essence$'$ and 
transforms it into the input format of a number of solvers (currently Minion~\cite{minion}, 
Gecode~\cite{gecode} and Dominion~\cite{dominion}). It was designed from the start to be solver-independent and
easily extended with new transformation rules. It is also straightforward to add
new output languages supported by an alternate sequence of transformations. 
At present Savile Row is at an early stage of development compared to 
other tools such as MiniZinc~\cite{minizinc}. However it has some features that are
particularly relevant to qualitative modelling, and its extensibility makes it 
suitable for the future work we describe below.

Uniquely Savile Row can produce Minion and Dominion's logical metaconstraints for conjunction 
and disjunction. This is highly relevant to qualitative modelling because
disjunctions arise from exists statements, and conjunctions from forAll statements
(when they are nested inside exists or some logical operator). Exists and forAll
will be extensively used in qualitative modelling to model time. Minion's logical 
metaconstraints can be much more efficient than other methods~\cite{watched-or-aij}.

Savile Row also implements common subexpression elimination (CSE) \cite{rendl-cse}. This replaces two
or more identical expressions in a model with a single auxiliary variable. The
auxiliary variable is then constrained to be equal to the common expression. 
In many cases CSE will strengthen propagation. CSEs tend to arise when quantifiers
are unrolled, so we expect this feature to be very relevant to QM. At present Savile Row
will only exploit identical common subexpressions. To fully exploit CSE for QM,
we would need to identify the types of non-identical CSEs that occur with QM 
(for example, common subsets of disjunctions) and extend Savile Row to eliminate
them. 

To better express complex QM problems in Essence$'$ is likely to require extensions to
the language. In particular we have identified comprehensions  
as an interesting future direction. These allow more flexible expression of 
constraints with respect to quantifier variables and parameters. For example, 
suppose we have a one-dimensional matrix $x$ and we want to state that there
exists a mid-point such that all variables before the mid-point are different, and 
the mid-point equals some parameter $p$. Using a variable comprehension, we can 
express this as follows. The comprehension creates a list of variables for the 
allDiff constraint. 

\begin{center}
    exists $i : int(0..max)$. allDiff($[x[j]\: |\: j : int(0..max),\: j<i ]$) $\wedge x[i]=p$
\end{center}

Comprehensions afford a great deal of flexibility. As a second example, they would 
allow the tuple lists of table constraints to be 
constructed on the fly based on parameters and quantifier variables. Therefore 
we expect them to be an excellent addition to the language for QM and for many
other problems. 

\subsection{Solver Generation}
\label{solvgen}

A major challenge facing constraints research is to deliver constraint solving
that scales easily to problems of practical size. Current constraint solvers,
such as Choco~\cite{chocosolver}, Eclipse~\cite{eclipsesolver}, Gecode~\cite{gecode}, Ilog
Solver~\cite{ilog}, or Minion~\cite{minion} are monolithic in design, accepting
a broad range of models. This convenience comes at the price of a necessarily
complex internal architecture, resulting in significant overheads and inhibiting
efficiency and scalability. Each solver may thus incorporate a large number of
features, many of which will not be required for most constraint problems. The
complexity of current solvers also means that it is often prohibitively
difficult to incorporate new techniques as they appear in the literature. A
further drawback is that current solvers perform little or no analysis of an
input model and the features of an individual model cannot be exploited to
produce a more efficient solving process.

To mitigate these drawbacks, constraint solvers often allow manual tuning of the solving process. However, this requires considerable expertise, preventing the widespread adoption of constraints as a technique for solving the most challenging combinatorial problems. The components of a constraint solver are also usually tightly coupled, with complex restrictions on how they may be linked together, making automated generation of different solvers difficult.

We address these challenges in the Constraint Solver Synthesiser project. The benefits achieved in the framework lead to faster and more scalable solvers. In addition, the automated approach simplifies the task of modelling constraint problems by removing the need to manually optimise specifications. As well as architecture-driven development, we utilise concepts from generative
programming, AI, domain-specific software engineering and product-lines in the
Constraint Solver Synthesiser approach. 

Initial results from comparing solvers generated by Dominion with an existing
solver are positive and indicate this approach is
promising~\cite{balasubramaniam_automated_2012}. Dominion is in fact expected to make bigger gains in the cases where there are many interdependent decisions to be made from a large number of components, where traditional solvers are limited by having to cater for the generic problem.

The Dominion approach improves performance and scalability of solving constraint problems as a result of:
\begin{itemize}
\item{tuning the solver to characteristics of the problem}
\item{making more informed choices by analysing the input model}
\item{specialising the solver by only incorporating required components, and}
\item{providing extra functionality that can be added easily and used when required.}
\end{itemize}

A number of avenues are open for further work. In particular learning how to
automatically create high quality solvers quickly is a major open problem. This
is essentially an instance of the Algorithm Selection
Problem~\cite{rice_algorithm_1976}. A lot of research has investigated ways of
tackling this problem, but veritable challenges remain. A prime example for new
challenges in Algorithm Selection are the issues related to contemporary machine
architectures with a large number of computing elements with diverse
capabilities (e.g.\ multiple CPU and GPU cores in modern laptops). Research to
date has largely focussed on using a single processor, with some research into
parallelisation on homogeneous hardware. Being able to run several algorithms at
once has a significant impact on how algorithms should be selected. In
particular, constraints on the type of algorithms that be run at the same time,
for example because only one of them can use the GPU, as well as collaboration
between the algorithms pose promising directions for research.

All of these directions are highly relevant to qualitative modelling, as
advances that speed up constraint solving in practise would enable us to tackle
practical problems that are currently beyond the reach of CP.

\subsection{Exploring Search Spaces I}
\label{space1}

Current CP solvers are tailored towards finding a single solution to a problem, or proving no solution exists. The solution found can be either the first one discovered, or the ``best'' solution under a single optimisation condition. In many situations this is insufficient, as users want to be able to understand and reason about \textit{all} solutions to their problem. For many such problems, current CP is simply useless. We believe CP solvers must be extended to be able to solve such problems, while maintaining and improving the efficiency and ease-of-use of existing CP tools.

Groups are one of the most fundamental mathematical concepts, and problems whose solutions are a group occur in huge numbers of both research and real-world applications. All groups include an ``identity'' element, so the problem of finding a single solution to a problem whose solutions form a group is trivial. Enumerating all solutions to such problems is impractical, as groups considered ``small'' by mathematicians often have over \(10^{100}\) elements.

The reason groups with more than \(10^{100}\) members can be handled is that groups are rarely represented by a complete enumeration. Instead, groups are represented by a small subset of their elements, which can be used to \textit{generate} the whole group, utilising the fact that groups are closed under composition of their members. Using a small number of members of a structure to generate the complete structure occurs in many areas of mathematics, including algebraic structures such as groups, semigroups, vector spaces and lattices. 

We plan on extending CP so it can generate efficient compact representations of the solutions to problems, and allow users to explore and understand these solutions. This will allow CP to be used to tackle many new classes of problems, of interest to many different types of user. 

A related issue is the parallel exploration of search spaces. This is an
especially relevant issue as during the last few years, a dramatic paradigm
shift from ever faster processors to an ever increasing number of processors and
processing elements has occurred. Even basic contemporary machines have several
generic processing elements and specialised chips for e.g.\ graphics processing.

While many systems for parallel constraint solving have been developed, we are
not aware of any in current use that can be deployed easily by non-expert users.
Recent work at St Andrews started to address this
problem~\cite{kotthoff_distributed_2010} and the latest released version of the
Minion constraint solver (version 0.14, July 2012) has preliminary support for
the large-scale distributed solving of any constraint problem. However, further
research is required to make it easier to use and evaluate its usefulness for
qualitative models.

\subsection{Exploring Search Spaces II}
\label{space2}

In Section \ref{space1} we described issues to do with the efficient search of large solution spaces, which is clearly of fundamental importance for QM. However, even if efficiency is assured, there are two further problems to overcome if high quality QM is to be achieved. The first is the organisation of search in a controlled and stochastic way -- i.e. using the mathematical theory of probability to express and utilise the inherent degrees of uncertainty in which qualitative model solutions are likely to lead to ``good" quantitative models. Existing CP search heuristics allow the user to specify the order in which the variables and/or values are selected during search. This order can be randomised (implemented for example as the {\it -randomiseorder} and {\it -randomseed} heuristic options in Minion), but this is far from a fully stochastic exploration of the search space. Both BIOCHAM and BioPEPA  (described in Section \ref{qualold}) fully support iterative stochastic simulation allowing convergence to preferred numeric models.

\begin{figure}[!ht]
\begin{center}
\includegraphics[width=4.5in]{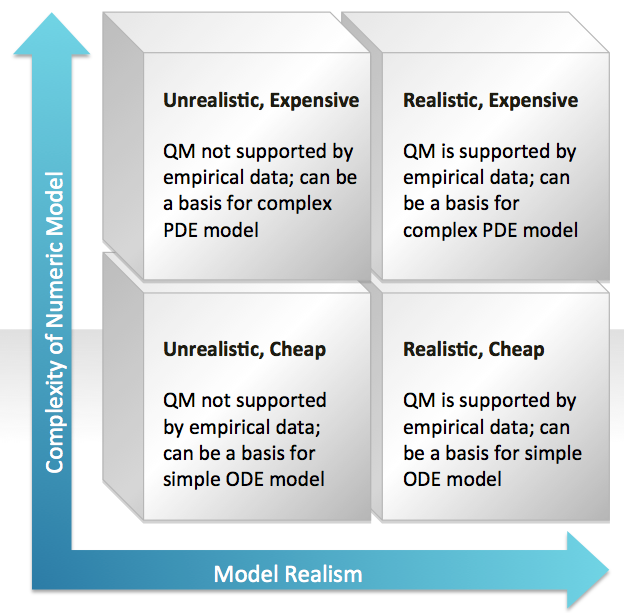}
\end{center}
\caption{Simplified tradeoff between QM realism and numeric model complexity}
\label{toff}
\end{figure}

The second issue relates to the tradeoff between scientific accuracy and plausibility of a QM (as determined by testing generalisation to empirical data) and the mathematical and computational complexity of the preferred quantitative model, as shown in Figure \ref{toff}. Qualitative models can be ranked in terms of realism in a continuum ranging from highly unrealistic to a highly accurate simulation of what we understand the system in question to be. The models can also be ranked in terms of the type of differential equations needed to implement a numeric simulation. Many simple systems of linear ODEs are solvable in polynomial time and space. Others are not (depending on Lipschitz conditions and whether or not $P = PSPACE$~\cite{Ko}). Nonlinear ODEs are strictly harder to solve as a class, and most PDEs have no closed form solution. The complexity of obtaining approximate solutions follows the same scale, in general. 
It is clear that given two qualitative models that are roughly equivalent in terms of assessed realism, the one that leads to the differential equations that are easier to solve should normally be selected. The CP technology needed to make these decisions does not exist, and its development is a completely unexplored avenue of future research.

\section{Conclusions}
\label{conc}

A large proportion of research effort in CP is directed inwards. Quite correctly, researchers seek ways to improve the modelling of CSPs, the efficiency of constraint propagators, and the range and scope of constraints in a general sense. This is as it should be, and the authors' combined research effort is predominantly inwards in this sense. However, if technologies such as CP are not being used by non-developers to solve problems in the wider domain, then they are of intellectual interest to only a small numbers of insiders. 

In this paper we describe an area of use for CP technologies that has fallen into neglect, in our opinion for no good reason. The temporal logic and  process calculus research communities are achieving success in qualitative modelling by publishing papers, being awarded grants, and by having the fruits of their research efforts used to solve real problems in systems biology. But dynamic systems can be perfectly well described in terms of finite difference relationships that obviate the need for temporal and process components in the underlying system description language. All finite difference methods rely on discretising a function on a grid, and the discretisation can be readily expressed in terms of CP variables and values with simple arithmetic constraints: in Section \ref{qualnew} we described the standard backward-difference approximation of a derivative, using unit step-length in order to maintain integer value  domains. Forward and central differences can be approximated using the same technique, as can derivatives to any required higher order. The fact that time is the dependent variable in our models is unimportant: the discretisation works for arbitrary choice of variable representation. That the numeric error in finite difference approximations of derivatives is proportional to the step size (one for our forward and backward differences; two for central differences) is also unimportant: our aim is to derive a CSP with larger than needed solution space, in order not to rule out realistic models that would not be result of {\it a priori} choice of differential equation model.

In addition, it is our view that the CP framework is inherently more attractive than temporal and process frameworks, since the ability to formally reason about a timeline in terms of ``until", ``since", etc. is not needed, and, if present, makes searching for solutions harder than necessary due to  well-documented problems with state-space explosion.

However, current CP technology is not well enough developed to compete with (and ideally replace) the areas of computer science that have dedicated more research effort and resource to this area of study. CP research effort into qualitative modelling faltered in the early years of this century, and has not yet recovered. The specific areas identified in Section \ref{future} are a non-exhaustive set of future research directions for the CP community that, if successful, would allow our languages and tools to be routinely used by researchers from the physical, biological and economic sciences.

\subsubsection*{Acknowledgments.} 
The authors are supported by United Kingdom  EPSRC grant EP/H004092/1. 
LK is supported by a SICSA studentship and an EPSRC fellowship.


\bibliographystyle{splncs03}      
\bibliography{QualCSP}   

\end{document}